\documentclass[sn-mathphys,Numbered]{sn-jnl}

\usepackage{graphicx}%
\usepackage{multirow}%
\usepackage{amsmath,amssymb,amsfonts}%
\usepackage{amsthm}%
\usepackage{mathrsfs}%
\usepackage[title]{appendix}%
\usepackage{xcolor}%
\usepackage{textcomp}%
\usepackage{manyfoot}%
\usepackage{booktabs}%
\usepackage{algorithm}%
\usepackage{algorithmicx}%
\usepackage{algpseudocode}%
\usepackage{listings}%
\usepackage{subfigure}%

\theoremstyle{thmstyleone}%
\theoremstyle{thmstyletwo}%

\theoremstyle{thmstylethree}%

\begin{document}

\title[Article Title]{Dynamic Fabry-P\'erot cavity stabilization technique for atom-cavity experiments}


\author*[1]{\fnm{S. P. } \sur{Dinesh}}\email{sreyaspd@rri.res.in}

\author[2]{\fnm{V. R. } \sur{Thakar}}

\author[3]{\fnm{V. I. } \sur{Gokul}}

\author[4]{\fnm{Arun } \sur{Bahuleyan}}
\author[6]{\fnm{S. A. } \sur{Rangwala}}

\affil*[1]{\orgdiv{Light and Matter Physics}, \orgname{Raman Research Institute}, \orgaddress{\street{Sadashivanagar}, \city{Bangalore}, \postcode{560080}, \state{Karnataka}, \country{India}}}


\abstract{We present a stabilization technique developed to lock and dynamically tune the resonant frequency of a moderate finesse Fabry-P\'erot (FP) cavity used in precision atom-cavity quantum electrodynamics (QED) experiments. Most experimental setups with active stabilization either operate at one fixed resonant frequency or use transfer cavities to achieve the ability to tune the resonant frequency of the cavity. In this work, we present a simple and cost-effective solution to actively stabilize an optical cavity while achieving a dynamic tuning range of over 100 MHz with a precision under 1 MHz.
Our unique scheme uses a reference laser locked to an electro-optic modulator (EOM) shifted saturation absorption spectroscopy (SAS) signal. The cavity is locked to the PDH error signal obtained from the dip in the reflected intensity of this reference laser.
Our setup provides the feature to efficiently tune the resonant frequency of the cavity by only changing the EOM drive without unlocking and re-locking either the reference laser or the cavity.
We present measurements of precision control of the resonant cavity frequency and vacuum Rabi splitting (VRS) to quantify the stability achieved and hence show that this technique is suitable for a variety of cavity QED experiments.}




\maketitle

\section{Introduction}\label{sec1}

Cavities find applications in quantum electrodynamics (QED) \cite{pur,jay,har,rem}, precision spectroscopy \cite{CRDS,spec1,spec2,jun} all the way to gravitational wave detection \cite{ligo1,ligo2}, to give a few diverse examples of the versatility of these systems. Several experiments study a multitude of physical phenomena originating from a combined system containing an optical cavity and an intra-cavity single atom or an ensemble of atoms \cite{vuletic,kaiser,ess,eitremp,yzhu,gold,magnus}. In these experiments, the optical cavity can be tuned to a desired frequency close to the atomic transition, and an ensemble of atoms can be cooled and confined to have spatial overlap with the mode defined by the geometric properties of the cavity, thus coupling the atoms to the cavity. In our experimental setup, we study the effects of interactions between ultra-cold dilute gas of atoms in a magneto-optical trap (MOT) and a resonant moderate finesse Fabry-Perot (FP) cavity \cite{tray,sdutta,rahul,niru} in which a weak probe beam along the cavity axis is typically scanned over several tens of megahertz within a few milliseconds. For effective measurements at low signal levels, the experiment needs multiple repetitions. To achieve this, the resonant condition of the cavity needs to be held constant with respect to the cavity mode coupled atoms. \par To perform these measurements efficiently, it is necessary to either actively or passively stabilize the FP cavity to compensate for any drifts in the length of the cavity due to temperature fluctuations, acoustic noise, etc. Passive stabilization is achieved by using ultralow thermal expansion materials and vibration-resistant mounting designs to build the cavity \cite{ule1, fox, uleyu, hirata, white, notcutt, baner} and active stabilization techniques use feedback loops to maintain the resonant frequency of the cavity \cite{active2,active1,active5, active4, active3}. The existing active stabilization techniques use transfer cavities to shift the lock point and hence achieve the dynamic tuning property. In this paper, we propose a Pound-Drever-Hall (PDH) \cite{PDH1, PDH2, PDH3} based active stabilization scheme of the cavity that uses only two lasers. The FP cavity is locked to a reference laser which is in turn locked to an electro-optic modulator (EOM) shifted, saturation absorption spectroscopy (SAS) signal. A feature of our setup allows us to tune the resonant frequency of the cavity over a range of over a hundred MHz with a sensitivity of 1MHz by introducing a shift in the reference laser frequency without unlocking either the cavity or the reference laser. Thus, a locking technique that can be dynamically tuned over a wide frequency range can be implemented in experiments where we want to study the dependence of light scattered by the coupled atom-cavity system as a function of the detuning between the atomic transition and cavity frequency. Our main objective in presenting this work is to show how one can develop a relatively low investment and robust lock setup that bypasses using the transfer cavity to achieve the dynamic tuning ability. In this paper, we describe, the general experimental setup, the technique implemented to tune and stabilize the cavity, the configuration of the reference laser, the setup used to generate the necessary error signal, and the sensitivity, effectiveness, and tuning range of the implemented technique.
\begin{figure}[t]
	\centering
		\includegraphics[width=\textwidth]{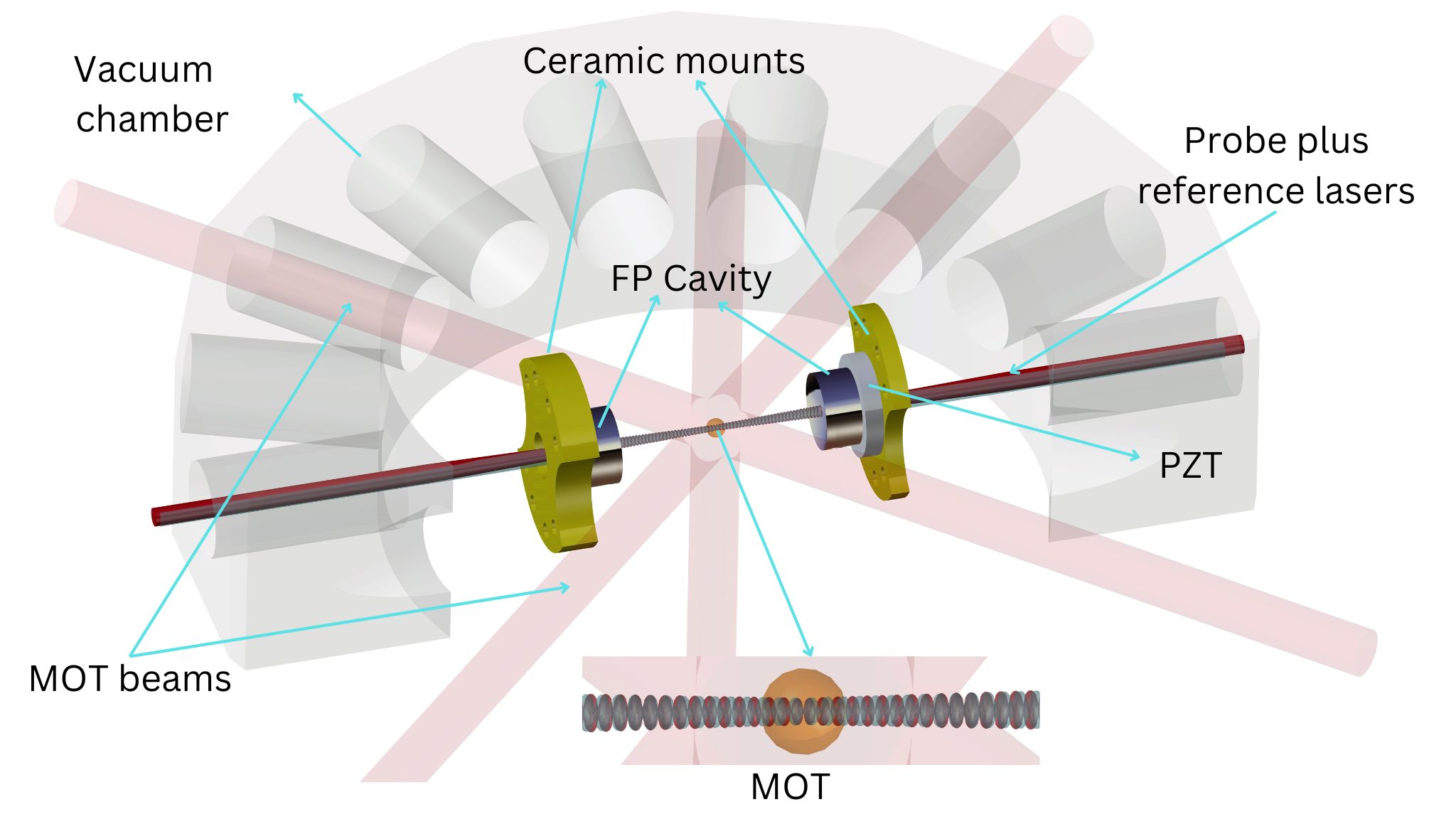}
		\caption{\textbf{Schematic of the experiment.} Six laser cooling beams (light brown) create the MOT between the cavity mirrors. The other element shown are annular piezoelectric transducer (PZT),  The description of the operation is in the text.}
  \label{MOT_setup}
\end{figure}
\section{Experimental setup}
The six laser cooling beams and a gradient magnetic field form the ${}^{85}$Rb MOT that confines atoms at the center of the FP cavity formed by two high reflectivity mirrors, as shown in Fig                                 
                       . \ref{cav_lock}. The Fabry-Perot cavity consists of two identical spherical mirrors of diameter ${12.5}$ mm, radius of curvature ${\approx}$ ${50}$ mm, separation of  ${\approx}$ ${45}$ mm, a beam waist of ${\approx}$  ${78}$  ${\mu}$m and has a linewidth of nearly ${\approx}$ ${8.8}$ MHz at 780nm. The input mirror is attached to an annular piezoelectric transducer (PZT). The 780 nm resonant probe beam (red) and the 767 nm cavity lock beam (blue) are coupled into the cavity so that they are both simultaneously resonant with the cavity mode. On the input side, a quarter-wave plate (QWP) and half-wave plate (HWP) combination in conjunction with two polarization beam splitting cubes (PBS) are used to separate the back reflected beams from the input mirror of the cavity, and the 767 light is captured by the photodiode (PD) to measure the light intensity dip in the reflected light when the cavity is resonant with the 767 nm as shown in \ref{cav_lock}. This allows the creation of a feedback signal to the PZT and locks the cavity length as discussed further in section B. 

The 780 nm probe light is transmitted through the appropriately tuned locked cavity for probing the atoms. The transmitted light through the cavity is passed through a premium 780 nm band pass filter (BPF) (Thorlabs-FBH780-10) and detected by the CCD camera and the photomultiplier tube (PMT). The spatial overlap of the cavity mode and the MOT density profile is measured to obtain the number of atoms coupled to the cavity. In our experiment, prominent collective strong coupling effects such as vacuum Rabi splitting (VRS) \cite{tavisvrs, agarwalvrs, rem} can be detected when a few thousands of atoms are coupled to the cavity. A ${780}$ nm laser referenced to a Rb saturated absorption spectroscopy (SAS) signal is used to probe VRS. This probe laser is incident along the axis of the cavity. As the atoms of interest are ${}^{85}$Rb, whose 5S\textsubscript{1/2} - 5P\textsubscript{3/2}(D2) transition wavelength is at ${780}$nm a separate wavelength has to be used to lock the cavity. This laser has to be far off-resonance from ${}^{85}$Rb D2 transitions to avoid any interaction with the cavity-coupled atoms. The coatings of the cavity restricted the choice of the reference laser to within ${780\pm}$100 nm. Keeping these conditions in mind, a ${767}$ nm laser is chosen as the reference laser to lock the cavity. A ${40}$ dB premium bandpass filter is placed in the cavity transmission collection setup to ensure that the contribution of the reference laser in the measured transmitted signal is well below the average noise level. The ${767}$ nm laser needs to be locked to a potassium SAS signal since the drift in its frequency changes the resonant frequency of the cavity and impacts the precision of the experiment This SAS locked frequency is noted using a wavemeter.

\subsection{Scheme of the lock}
Initially, the ${780}$ nm probe laser is locked to an independent rubidium SAS setup while the FP cavity and the ${767}$ nm reference laser are kept unlocked. A CCD camera is used to observe the spatial profile of the transmitted light from the cavity. The unlocked cavity is tuned to transmit TEM\textsubscript{00} mode of ${780}$ nm probe laser by adjusting the voltage of the cavity PZT. As the cavity frequency is scanned about this TEM\textsubscript{00} mode over a range of ${40}$ MHz, the characteristic Lorentzian peak is observed in the cavity transmission signal. Any major drifts or fluctuations in the cavity length are countered by manually adjusting the cavity PZT voltage in order to ensure that this transmission peak, corresponding to the resonance condition for TEM\textsubscript{00} mode of probe laser, would be seen in each scan cycle. While maintaining this transmission peak within the scan range, the frequency of the ${767}$ nm reference laser is changed by gradually varying the laser controller offset. This is done until a dip in the reflected intensity of the reference laser and the corresponding PDH error signal is observed to coincide with the transmission peak of the probe on an oscilloscope. Thus, a condition is experimentally achieved where the cavity can be stabilized using the PDH error signal produced by the ${767}$ nm reference laser while being simultaneously nearly resonant to the TEM\textsubscript{00} mode of the ${780}$ nm probe. At this point, the cavity length is approximately a common multiple of both the wavelengths corresponding to the probe and the reference laser. Let, ${L}$ be the cavity length, ${c}$ the speed of light, ${\nu}$\textsubscript{${p}$} and ${n}$\textsubscript{${p}$} respectively, be the frequency and the longitudinal eigen mode number corresponding to the TEM\textsubscript{00} mode of the probe laser sustained in the cavity, ${\nu}$\textsubscript{${r}$} and ${n}$\textsubscript{${r}$} be the frequency and longitudinal eigen mode number of the mode of reference laser. Then while keeping ${\nu}$\textsubscript{${p}$} and ${n}$\textsubscript{${p}$} constant, ${\nu}$\textsubscript{${r}$} and in turn, ${n}$\textsubscript{${r}$}  are varied until the condition    ${L}$ ${\approx}$ ${c}$ ${\cdot}$n\textsubscript{p}/${\nu}$\textsubscript{p} ${\approx}$ c${\cdot}$n\textsubscript{r}/${\nu}$\textsubscript{${r}$} is satisfied. The reference laser frequency corresponding to this condition is noted on the wavemeter. This frequency is different from the SAS locked value. This preliminary measurement allows an appropriate choice combination of EOM and acoustic-optic modulator (AOM) required to introduce the necessary shift in the reference laser frequency from its SAS locked value. The reference laser is stabilized at the approximately required frequency and the FP cavity is in turn locked to the corresponding PDH error signal. Finally, the reference laser frequency is fine-tuned, without unlocking either the cavity or the laser itself, by adjusting only the EOM drive to maximize the transmitted intensity of the probe, thus achieving the desired condition, L = ${c}$  ${\cdot}$n\textsubscript{${p}$}/${\nu}$\textsubscript{${p}$} = ${c}$ ${\cdot}$n\textsubscript{r}/${\nu}$\textsubscript{r}.
\begin{figure}[H]
	\centering
		\includegraphics[width=\textwidth]{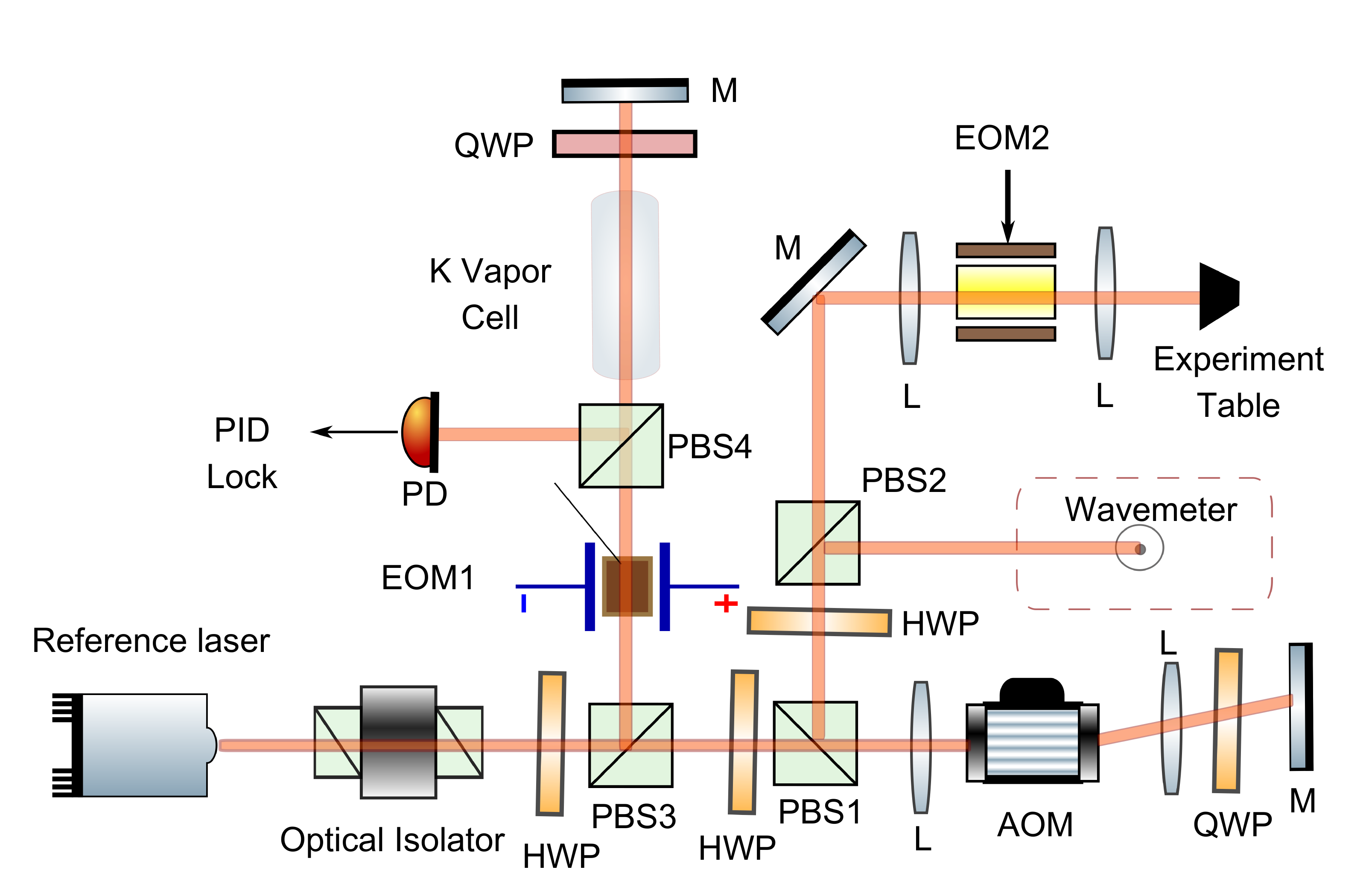}
\caption{\textbf{Setup of 767 nm reference laser}. Schematic representation of the optical circuit of the reference laser. Labels: M mirror, L lens, K potassium, HWP half-wave plate, QWP quarter wave plate, PBS1, PBS2, PBS3, PBS4 are polarizing beam splitters, AOM acoustic-optic modulator, EOM1 fiber electro-optic modulator, EOM2 free space electro-optic modulator.  The output of DL ${100}$ is passed through a combination of AOM and fiber-based EOM to obtain the desired shift in the frequency. Before the output is taken to the experimental table, it is passed through a free-space EOM to create sidebands for PDH locking.}
    \label{ref_laser}
\end{figure}
\subsection{Configuration of the reference laser}
A Toptica DL-100 extended cavity diode laser operated at ${767}$ nm, stabilized to potassium SAS is used as the reference to lock the cavity, see Fig. \ref{ref_laser} for the schematic of the experimental setup of the reference laser. The ${767}$ nm reference laser is transmitted through a fiber-based EOM to achieve the relative frequency shift between the probe and reference resonances, modulo the cavity-free spectral range (FSR). In an earlier attempt, the EOM output was fed to a potassium SAS setup to obtain three SAS signals. One for the central frequency and one for each sideband generated by the EOM. In this scheme, the ${767}$ nm laser is directly referenced to the SAS signal corresponding to the appropriate sideband. 

To avoid this issue in the present scheme the reference laser output from the Toptica DL-${100}$ is divided into two arms. The first arm is passed through a double pass AOM setup driven at ${200}$ MHz to generate a ${400}$ MHz shift in the reference laser frequency. The second arm is passed through the EOM to achieve the remainder of the shift. At this drive frequency, sufficient power can be transferred to the sideband, and a high-quality error signal for the sideband SAS can be obtained to lock the reference laser. After the double pass AOM setup, the first arm is further subdivided into two parts, one of which is fed to a wavemeter to monitor the reference laser frequency, and the other is passed through a free space Qubig EOM and coupled into a polarization maintaining single mode optical fiber to transfer it to the experimental table. When the cavity is tuned in resonance with the reference laser, the intensity of reference laser light reflected from the input mirror of the cavity dips.

The free space EOM is driven by the internal oscillator of the PDH module used to lock the cavity to generate the sidebands required to obtain the error signal from this reflection dip. On stabilizing the ${767}$ nm reference laser to the potassium SAS signal, the wavemeter readings showed a standard deviation of ${\approx}$ ${100}$ kHz, which is about the least count of the instrument. Since the cavity is made to follow the frequency of the reference laser, the length of the cavity is stabilized at a comparable frequency linewidth.

\begin{figure}[H]
	\centering
		\includegraphics[width=\textwidth]{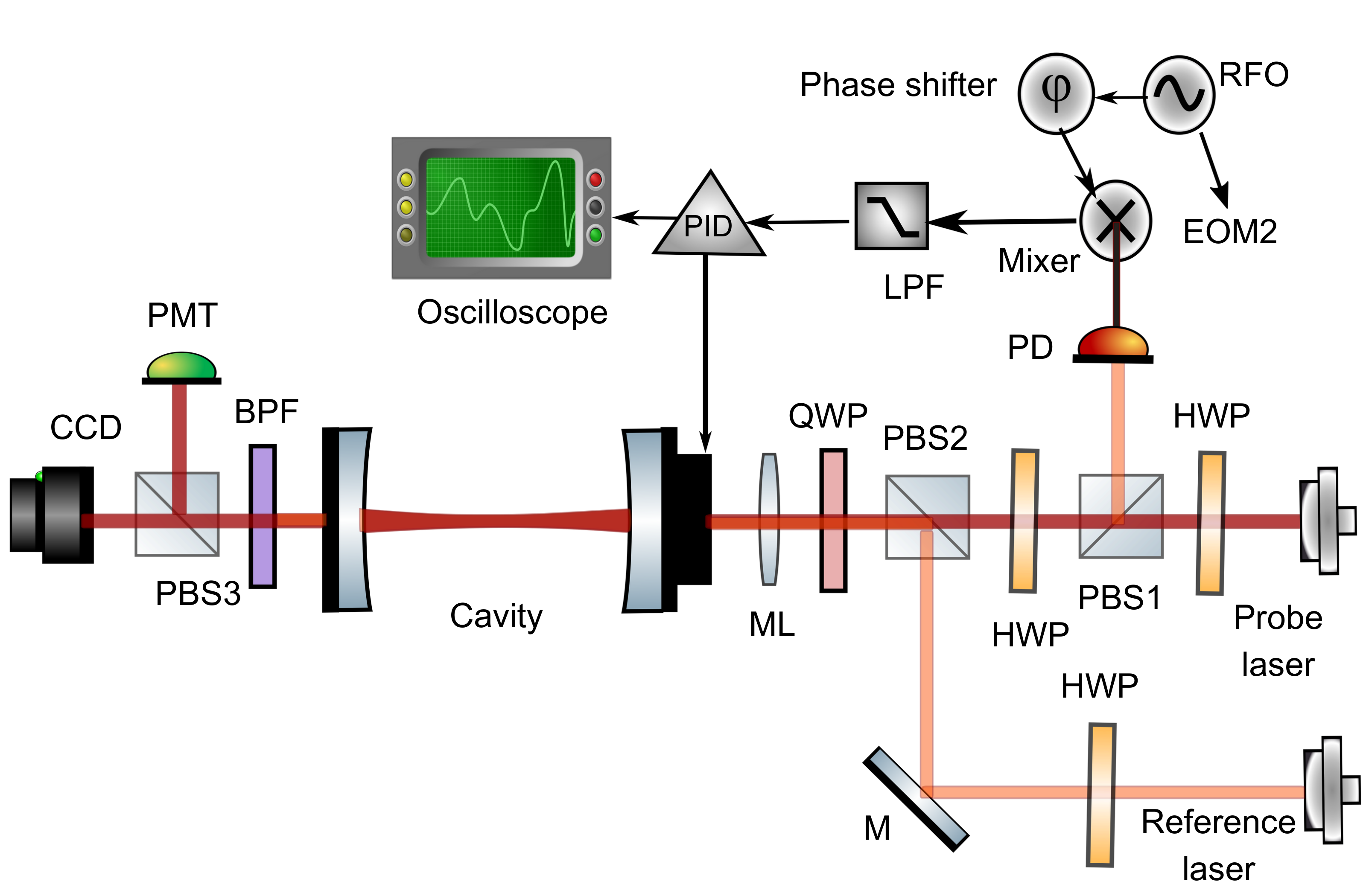}
		\caption{\textbf{Schematic of the experimental setup implemented to reference the cavity.} The stabilized reference laser is transferred to the experimental table through an optical fiber. Labels: QWP quarter waveplate, HWP half waveplate, PBS polarizing beam splitter, ML mode matching lens, HWP half-wave plate, QWP quarter waveplate, BPF premium bandpass filter, M mirror, CCD charge-coupled device, PD photodiode and PMT photomultiplier tube, RFO rf oscillator, LPF low pass filter, PBS1, PBS2 are polarizing beam splitters.}
  \label{cav_lock}
\end{figure}

\subsection{Cavity referencing setup}

The probe beam and the modulated reference laser are mixed using a half-wave plate and PBS2  as shown in Fig. \ref{cav_lock}. The output of the PBS is aligned along the axis of the cavity. A ${150}$ mm focal length convex lens is used to match the spatial mode profile of the two lasers with that defined by the FP cavity. This lens ensured that the two lasers are maximally coupled to the cavity. A quarter-wave plate is placed close to the window of the vacuum chamber. This quarter-wave plate introduced the shift in polarization required to separate out and measure the reflected intensity of the reference laser on a photodiode. On resonance, the cavity transmits the reference laser and a corresponding dip can be observed in the photodiode signal. The reflective dip is the order of ${1\%}$ dip in the reflected intensity. 

\begin{figure}[H]
	\centering
		\includegraphics[width=0.75\textwidth]{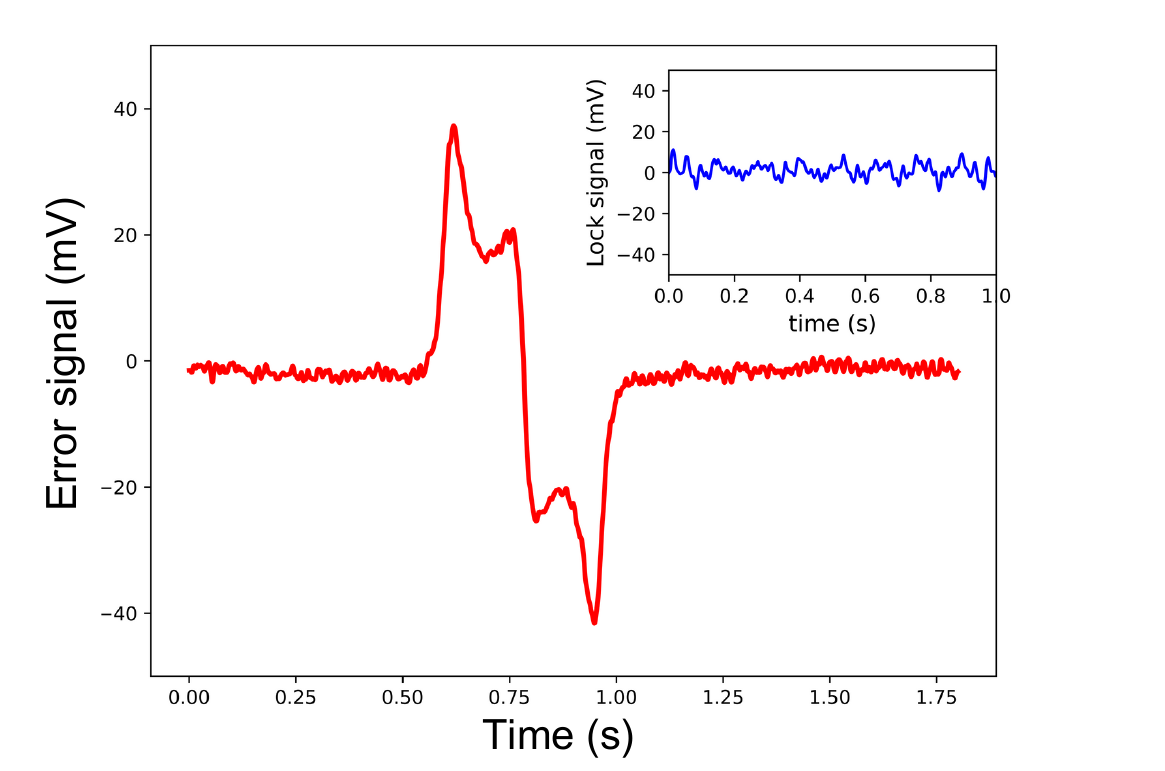}
		\caption{\textbf{PDH error signal to lock the cavity. Inset is the oscilloscope trace of a locked signal.} Cavity is scanned across the resonance of the ${767}$ nm reference laser to obtain the standard PDH error signal (red curve) from the dip in the reflected intensity. The inset graph (blue curve) shows the oscilloscope trace of a locked signal.}
    \label{Error_sig}
\end{figure}

The PDH internal oscillator drove the free space EOM to modulate the laser frequency at ${20}$ MHz to produce sidebands in the reflection dip. Since the cavity linewidth at ${767}$ nm is over ${10}$ MHz, the two sidebands can not be well resolved from the central feature in the dip. The shape of the error signal can be adjusted to produce a steep slope near resonance by carefully choosing the phase. At low oscillator power, the error signal strength increased with an increase in the amplitude of the PDH internal oscillator. However, as the sidebands are not well separated, a further increase in the amplitude of the internal oscillator would distort the signal. Despite the small amplitude and limited resolution of the dip, a good error signal is generated by optimizing the amplitude and the phase of the PDH internal oscillator, see Fig. \ref{Error_sig}. The optimized error signal is fed to a proportional, integration and derivative (PID) module which controls the cavity PZT voltage, thus stabilizing the cavity length. Fig. \ref{Error_sig} shows the optimized PDH error signal and the PID signal upon locking the cavity. The ${780}$ nm bandpass filter in the collection setup ensured that of both the frequencies transmitted by the cavity, only the contribution from ${780}$ nm probe is measured by the photo-multiplier tube (PMT).

\begin{figure}[H]
	\centering
		\includegraphics[width=0.65\textwidth]{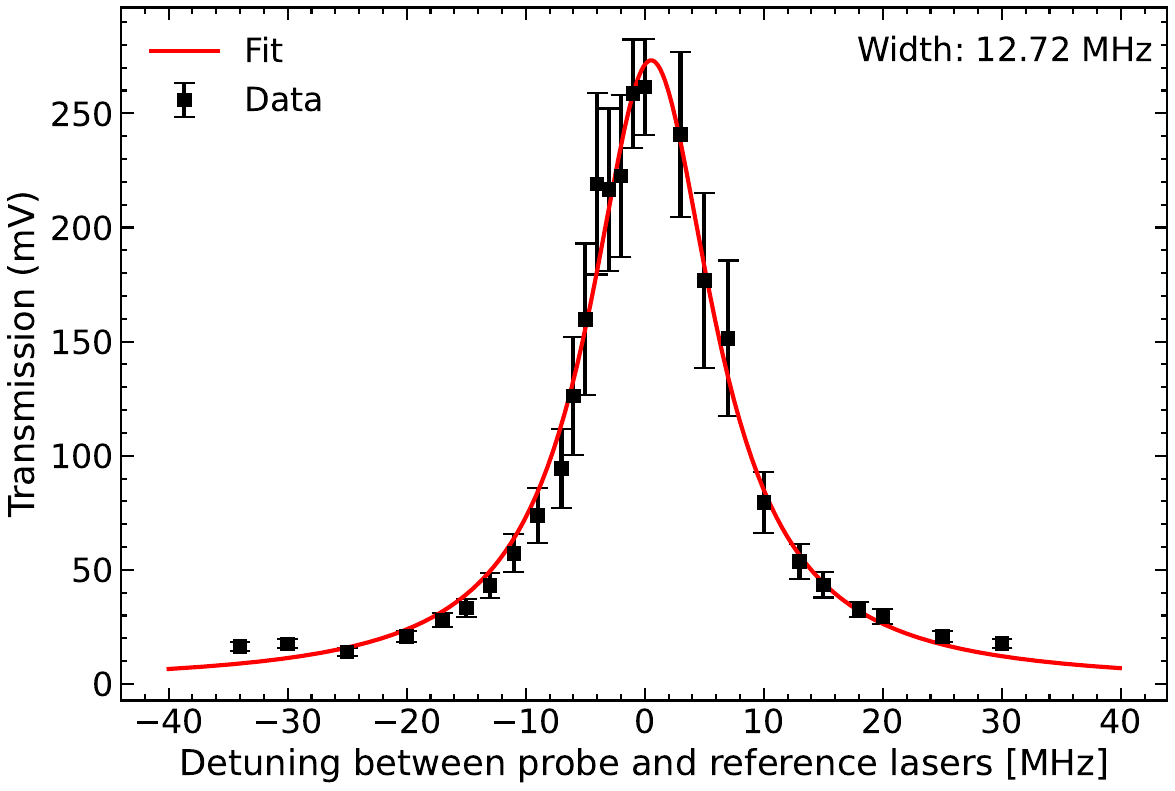}
		\caption{\textbf{Transmitted intensity of probe versus reference laser detuning.} The plot shows the intensity of transmitted ${780}$ nm probe light from the cavity versus wavemeter reading reference laser frequency. In this measurement, the cavity is locked to the reference laser, and the transmission of the probe is measured for various detunings of the reference laser for approximately two seconds. Each black dot represents the mean and the error bar is the standard deviation of the transmitted intensity at a given detuning. The value of the error bar is observed to increase near zero detuning. This is because of the  Lorentzian nature of the transmission of light through an FP cavity. The reference laser frequency is changed in steps and the intensity of the transmitted probe light is measured.}
  \label{linewidth}
\end{figure}

\section{Sensitivity and tuning ability of the lock}
Following the extensive setup, an experiment is conducted to measure the intensity of the probe transmitted through the empty cavity as a function of reference laser frequency. The probe laser is locked to F=${3}$ to F$'$=${3}$ D2 line of ${}^{85}$Rb transition. The reference laser is locked to the appropriate sideband SAS signal and the cavity is locked to the error signal of the reflection dip. The drive frequency of the fiber-based EOM (Eospace) is changed in steps of ${1}$ MHz by monitoring the wavemeter reading of the reference laser frequency. This effectively changes the resonant frequency of the cavity in steps of ${1}$ MHz. It must be noted that since $\Delta\nu_{p}=\Delta\nu_{r}\cdot n_p/n_r$, where $n_p/n_r\ne1 $, a change of ${1}$ MHz in ${767}$ nm reference laser frequency will not translate to ${1}$ MHz change in the resonant frequency of the ${780}$ nm probe. The intensity of the TEM\textsubscript{00} mode of the ${780}$ nm probe transmitted by the cavity is measured at each step for 10 seconds. The graph in Fig. \ref{linewidth} shows the average intensity and the standard deviation versus the resonant frequency of the cavity. The ability to distinguish between the average intensity in steps of ${1}$ MHz and the flexibility to tune the cavity over a range ${100}$ MHz without unlocking and relocking any of the components demonstrates the precision, sensitivity, and tunability of the cavity lock. 

To demonstrate the utility of the cavity stabilization mechanism in cavity QED experiments, we present a comparison of symmetry and precision in measuring VRS in our atom-cavity experiment before and after the implementation of the locking scheme. In the collective strong coupling regime, the split between the two normal modes of VRS is given by $g_0\sqrt{N_c}$, where $g_0$ is the atom-cavity interaction strength for a single atom and $N_c$ is the effective number of atoms coupled to the cavity.

We observe a double peak structure characteristic of VRS on probing at F=${3}$ to F$'$=${3}$ of  ${}^{85}$Rb D2 by using a weak light incident along the axis of the cavity \cite{tray}. The F=${3}$ to F$'$=${4}$ of ${}^{85}$Rb D2 laser is used for cooling transition for our MOT and needs to be turned off to ensure that the maximum number of atoms are prepared at the ground state of the probing transition. The MOT isotropically expands with a time constant of the order of a few milliseconds. Therefore, the probe is scanned within a millisecond, demanding a minimum scan rate of the order of ${1}$ kHz. This demands a higher bandwidth PMT pre-amplifier. In the higher bandwidth setting, the pre-amplifier generates higher noise, thus reducing the signal-to-noise ratio (SNR) in the transmission. To improve the SNR, the transmitted signal has to be averaged over several cycles. Before implementing the cavity stabilizing mechanism, small drifts, and fluctuations in the cavity length during the several consecutive cycles, would make the ${2}$ peaks in the VRS asymmetric as seen in Fig. 6a, 6b, and 6c. Once the cavity is stabilized using the reference laser, an average of over ${100}$ cycles can be successfully performed and we get a symmetric VRS as shown in Fig. 6d, which can be used to estimate the number of atoms coupled to the cavity with much higher reliability.

\begin{figure}[H]
	\centering
		\includegraphics[scale=0.7]{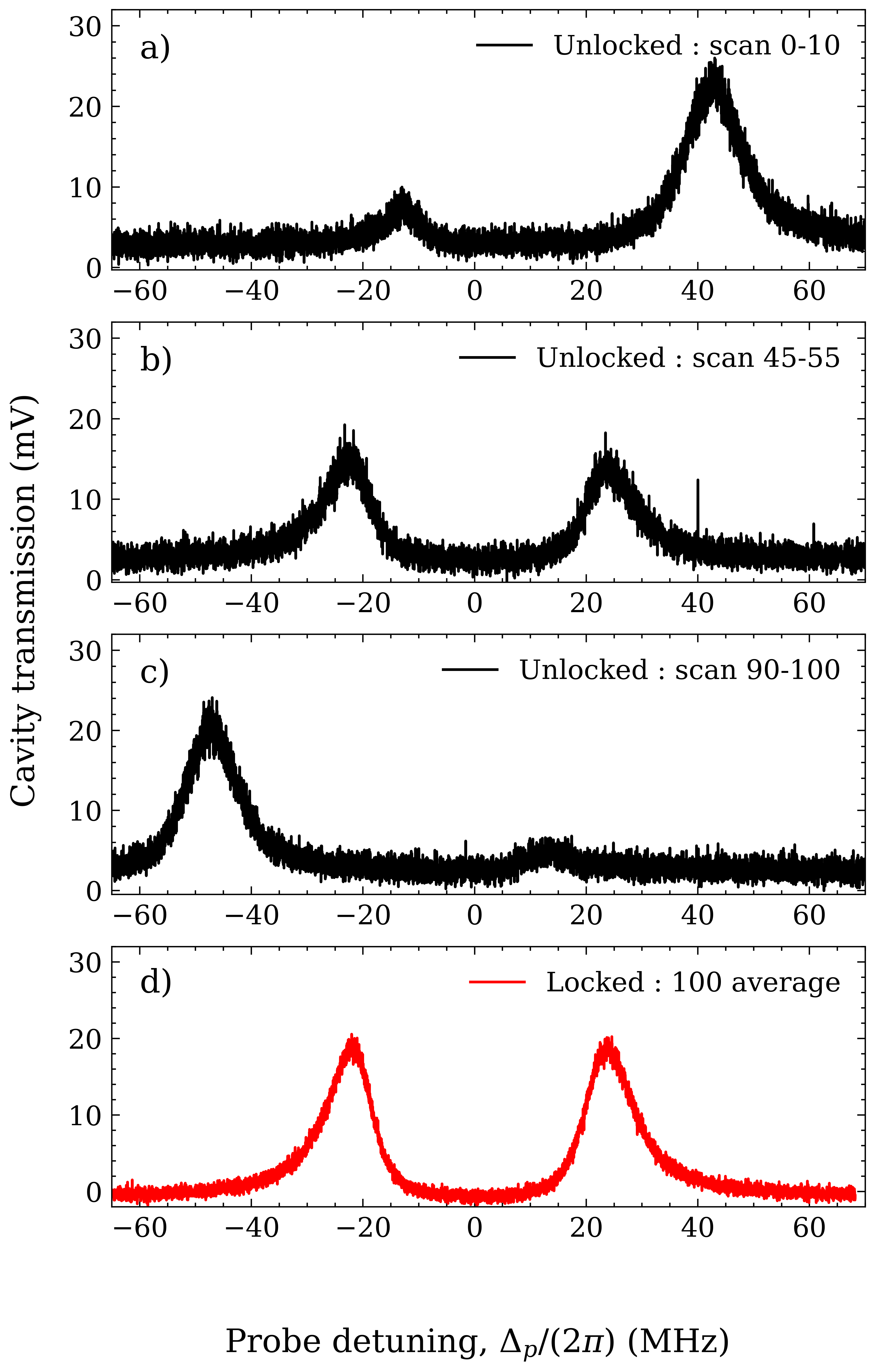}
		\caption{\textbf{VRS of a locked versus an unlocked cavity} The plot shows the averaged transmitted intensity of the probe versus detuning of the probe light from F=${3}$-${3}$$'$ transition of ${}^{85}$Rb under collective strong coupling condition. The two peak structure VRS is obtained by scanning the probe laser around the atomic transition. Fig (a), Fig (b), and Fig (c) are plotted by taking the averages of the data from $0-10$, $45-55$, and $90-100$ respectively. The symmetric VRS peak shown in Fig (d) is obtained when there is an active stabilization on the length of the cavity. In the unlocked case, during the measurement, the resonant frequency of the optical cavity changes due to the drift and fluctuations in the length, as a result, the expected symmetry in the double-peak structured VRS is lost.}
  
\end{figure}

\begin{figure}[H]
    \centering
    \subfigure[]{\includegraphics[width=0.48\textwidth]{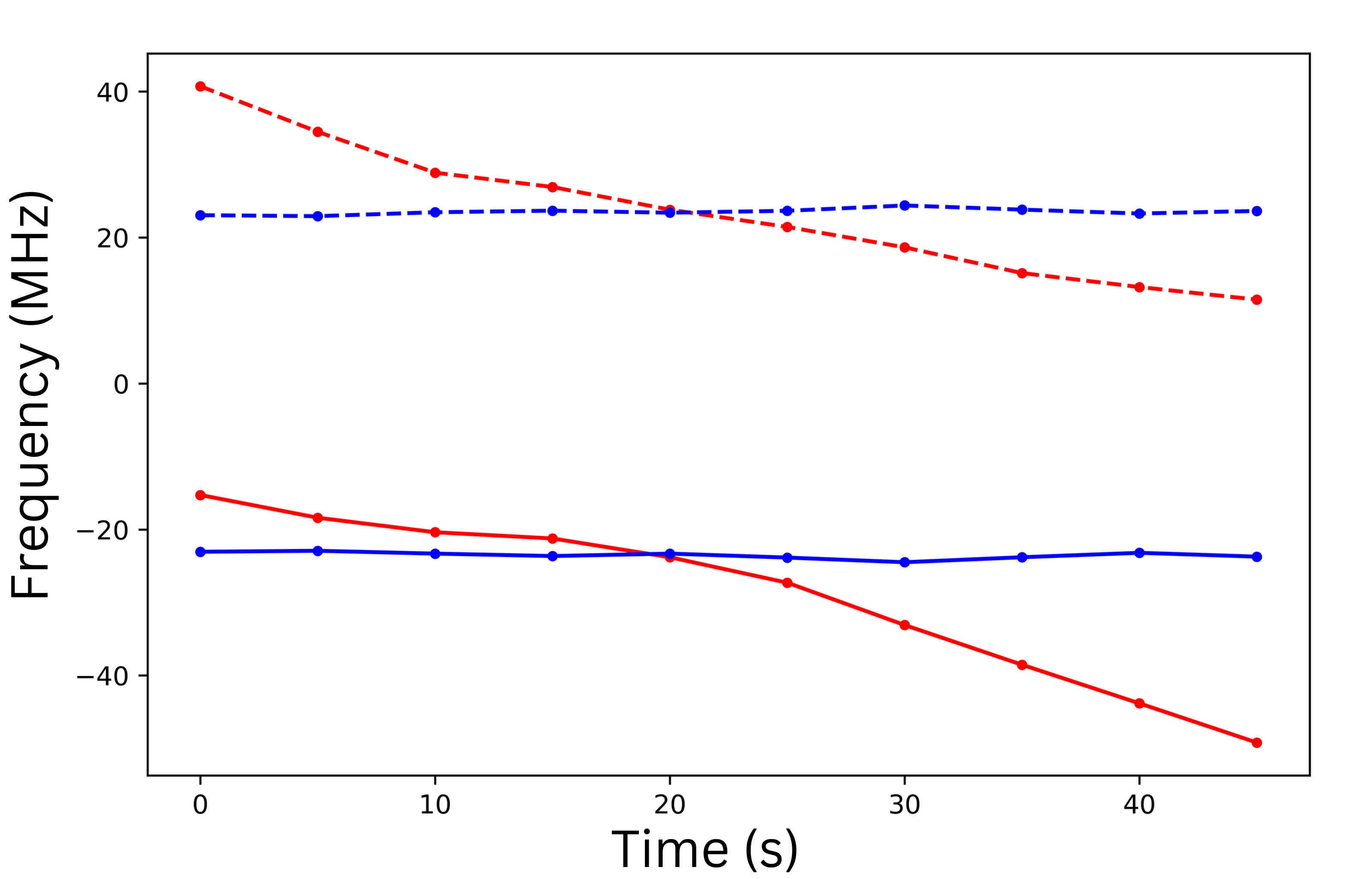}}
    \centering
    \subfigure[]{\includegraphics[width=0.47\textwidth]{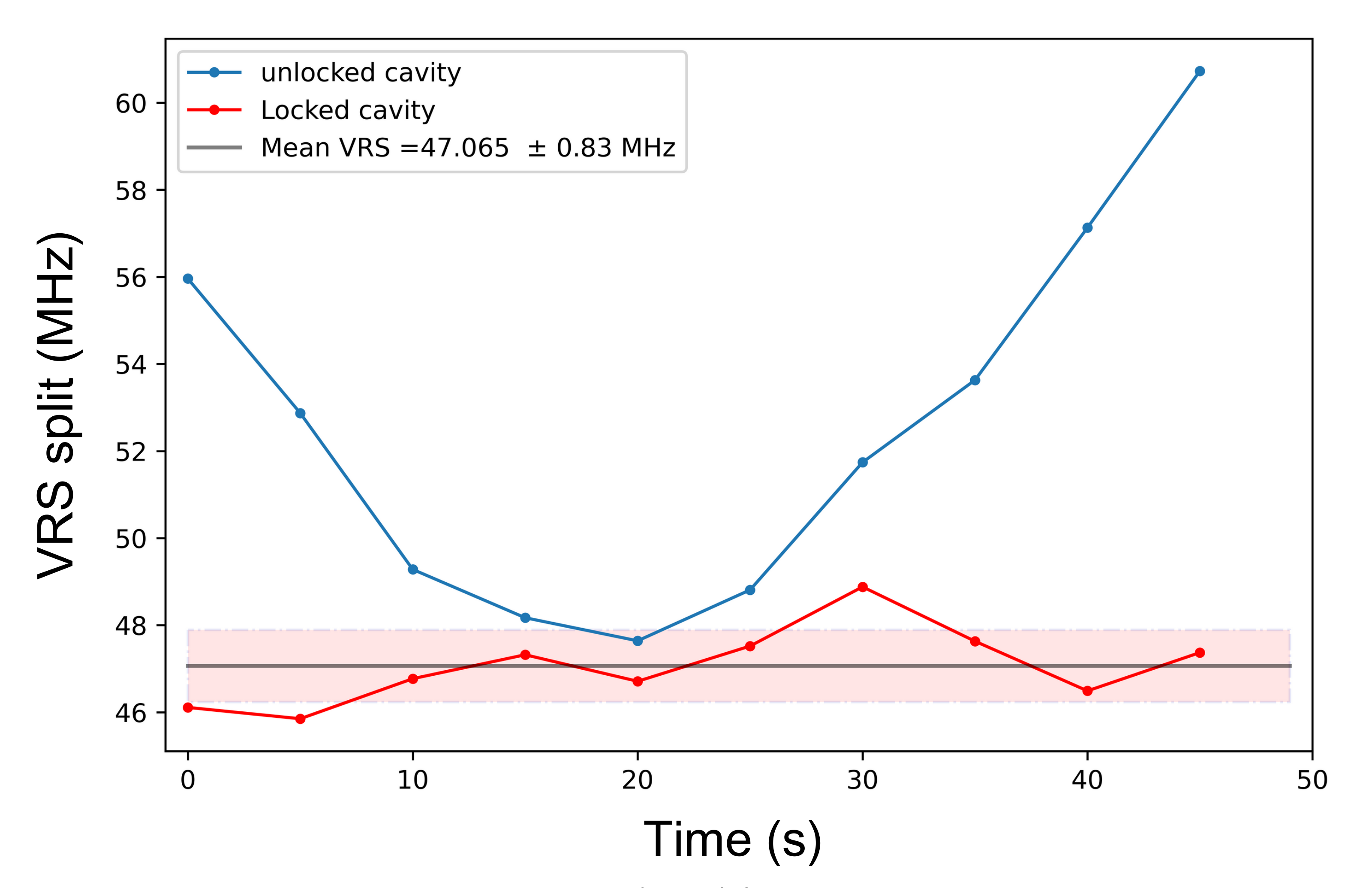}}
    
    \caption{\textbf{The plot (a) shows the drift of the unlocked cavity with time and (b) is the measured VRS split of an unlocked cavity versus a  locked cavity.} We allowed the cavity to drift and the probe laser is continuously scanned across the resonant transition to obtain the VRS split at various times, each dot is data of the VRS split averaged over 10 such scans. In the unlocked case, the fluctuations in the length of the cavity cause changes in resonant frequency and this introduces an unequal detuning between the atom and the cavity during the measurement cycle. In Fig (7a) we show the drift of the left and right peaks of the double peak structured VRS, the solid red plot shows the progression of the left peak without the lock, and the solid blue is the left peak with the cavity lock, the dotted red and blue shows the progression of the right peak without and with the cavity lock respectively. It can be observed that measured drifts in frequency for blue lines do not change with time. In Fig (7b) we show the VRS split observed for unlocked (blue plot) and locked (red plot).  A consistent VRS split with a standard deviation of $0.83$ MHz, when there is an active lock on the cavity is in place. The red dots are experimental data, The black solid line is the mean, and the light red band shows the standard deviation.} 
    \label{fig:ab}
\end{figure}
It can be clearly observed that the split in the VRS changes with time in the absence of the cavity lock due to cavity drifts, while it remains unchanged with the cavity lock in place as shown in Fig. 7. This demonstrates the efficiency of our FP cavity stabilization technique.

\section{Conclusion}
In this paper, we have discussed in detail the experimental setup and the techniques used to stabilize a Fabry-Perot cavity. We have demonstrated the ability to dynamically tune the resonant cavity frequency over a range of ${100}$ MHz with under ${1}$ MHz precision. The techniques present here are suitable for a variety of cavity QED experiments. They are especially applicable in studying the interaction between a detuned cavity and ultra-cold dilute gas of atoms \cite{rahul}. Such a lock strategy can be made compact and portable for future quantum electrodynamics experiments.

\section*{Acknowledgments}
The authors would like to thank the Department of Science and Technology and  Ministry of Electronics and Information
Technology (MeitY), Government of India, under a Centre for Excellence in Quantum Technologies grant with Ref. No.
4(7)/2020-ITEA. The authors would also like to acknowledge the inputs from Dr. Arijit Sharma, Mohamed Ibrahim, and Meena M S. 



\bibliography{main}


\end{document}